# The gravitomagnetic field of a sphere, Gravity Probe B and the LAGEOS satellites

Jacob Biemond*

*Vrije Universiteit, Amsterdam, Section: Nuclear magnetic resonance, 1971-1975*
*Postal address: Sansovinostraat 28, 5624 JX Eindhoven, The Netherlands*
Website: http://www.gravito.nl/ Email: j.biemond@gravito.nl

## ABSTRACT

The gravitomagnetic field generated by a rotating sphere is usually calculated from the ideal dipole model. However, for a sphere with a homogeneous mass density, this model is not generally valid. Trying to obtain a more accurate value of the gravitomagnetic field inside and outside the sphere, series expansions for this field are presented in this paper. The calculated polar gravitomagnetic field of the sphere and that from the ideal dipole model appear to coincide, but the field in the vicinity of the sphere may deviate. The deduced field within the sphere strongly deviates from the ideal dipole result.

As an illustration, the gravitomagnetic precession rate (or frame-dragging effect) of a gyroscope moving in the gravitomagnetic field from a large rotating sphere is calculated. For the Gravity Probe B experiment the result may coincide with the prediction from the ideal dipole model and in fair agreement with observations. In addition, the obtained Lense-Thirring precession rate for the LAGEOS satellites probably coincides with the standard prediction. For both experiments alternative predictions are calculated, when the gravitomagnetic field and the magnetic field from moving charge are equivalent. Theoretical and observational indications for such an equivalence are summarized.

The obtained series expansions for the gravitomagnetic field of a sphere can also be applied to the calculation of the magnetic field, generated by a rotating sphere with a homogeneous charge density. Results for this case are also discussed.

## 1. INTRODUCTION

The striking analogy between the magnetic field generated by moving charge and a so-called "magnetic-type" gravitational field generated by moving mass has been noticed for a long time. Heaviside (see, e.g. [1]) already derived Maxwell-type gravitational equations. Since that time more definitions of a gravitomagnetic field and alternative deductions of gravitomagnetic equations, analogous to the Maxwell equations, have been proposed by several authors (see for a review, e.g., [2]). Deductions of the gravitomagnetic equations from the Einstein equations in the slow motion and weak field approximation have been given by Peng [3], Biemond [2, 4, 5], Mashhoon [6], Pascual-Sánchez [7], Ruggiero and Tartaglia [8] and many others. The concept of a gravitomagnetic field generated by a spinning mass has more systematically been introduced later on in the development of the theory of general relativity [2, 4, 5]. Usually, its effects are described in terms like frame-dragging, but the concept of a gravitomagnetic field may give more insight. Therefore, we will exclusively apply the latter approach in this work.

In the stationary case the gravitomagnetic field **B** can be obtained from the simplified gravitomagnetic equations [2, 4, 5]

$$\nabla \times \mathbf{B} = -4\pi \beta c^{-1} G^{\frac{1}{2}} \rho \mathbf{v} \quad \text{and} \quad \nabla \cdot \mathbf{B} = 0, \qquad (1.1)$$

where **v** is velocity and $\rho$ is the density of a mass element $dm = \rho\, dV$. Choosing a dimensionless constant $\beta$ in (1.1), the field **B** obtains the dimension of a magnetic induction field. Depending on the definition of the field **B**, different values for the dimensionless constant $\beta$ of order unity have been introduced in the past (see ref. [5, p.7] for a discussion of this point). Moreover, alternative choices for the dimension of the field **B** are mathematically possible, leading to other dimensions for $\beta$. Furthermore, it is noticed that Gaussian units are used throughout this paper.

Since $\nabla \cdot \mathbf{B} = 0$, the field **B** can be derived from a gravitomagnetic vector potential **A** ($\mathbf{B} = \nabla \times \mathbf{A}$). The field **B** can be derived in a similar way as in the corresponding electromagnetic case (see, e.g., Landau and Lifshitz [9, § 43 and § 44])). For a massive rotating sphere with angular momentum **S** the following expression satisfies to (1.1)

$$\mathbf{A} = \tfrac{1}{2} \beta c^{-1} G^{\frac{1}{2}} \mathbf{S} \times \nabla \left( \tfrac{1}{R} \right). \tag{1.2}$$

In deriving (1.2) it has been assumed that the distance $R$ from the centre of the sphere is large compared with the radius $r_0$ of the sphere. In addition, the angular momentum **S** of the sphere with total mass $m$ is given by

$$\mathbf{S} = I\boldsymbol{\omega}, \text{ or } S = I\omega = \tfrac{2}{5} f_s m r_0^2 \omega, \tag{1.3}$$

where $\omega = 2\pi\nu$ is the angular velocity of the sphere ($\nu$ is its rotational frequency), $I$ is the moment of inertia of the sphere and $f_s$ is a dimensionless factor depending on the homogeneity of the mass density $\rho$ of the sphere. For a homogeneous mass density $f_s = 1$, but when if the mass density is greater near the centre of the sphere $f_s$ will be less than unity value.

From (1.2) the following expression for the gravitomagnetic field **B** follows

$$\mathbf{B} = -\tfrac{1}{2} \beta c^{-1} G^{\frac{1}{2}} \nabla \left\{ \mathbf{S} \cdot \nabla \left( \tfrac{1}{R} \right) \right\}. \tag{1.4}$$

This expression represents the gravitomagnetic field **B** of an ideal dipole, located in the centre of the sphere. The gravitomagnetic dipole moment **M** can be written as

$$\mathbf{M} = -\tfrac{1}{2} \beta c^{-1} G^{\frac{1}{2}} \mathbf{S}. \tag{1.5}$$

Combination of (1.4) and (1.5) leads to the following formula for the gravitomagnetic field

$$\mathbf{B} = \nabla \left( \mathbf{M} \cdot \nabla \frac{1}{R} \right) = \left( \frac{3\mathbf{M} \cdot \mathbf{R}}{R^5} \mathbf{R} - \frac{\mathbf{M}}{R^3} \right), \tag{1.6}$$

where $R$ is the distance from the centre of the sphere to the field point where **B** is measured ($R \geq r_0$). Note that this equation is completely analogous to the corresponding electromagnetic expression. In case of a homogeneous mass density ($f_s = 1$), combination of (1.3), (1.5) and (1.6) then yields the following gravitomagnetic fields **B** at the poles and at the equator of the sphere, respectively

$$\mathbf{B}_p(\text{sphere}) = -\tfrac{2}{5} \beta c^{-1} G^{\frac{1}{2}} m r_0^2 R^{-3} \boldsymbol{\omega}, \quad \mathbf{B}_{eq}(\text{sphere}) = \tfrac{1}{5} \beta c^{-1} G^{\frac{1}{2}} m r_0^2 R^{-3} \boldsymbol{\omega}. \tag{1.7}$$

The fields $\mathbf{B}_p(\text{sphere})$ and $\mathbf{B}_{eq}(\text{sphere})$ have been calculated under the assumption that $R \gg r_0$, so that these results may be only approximately valid in the vicinity of the sphere. Furthermore, it is noticed that neither the sign nor the value of $\beta$ follow from the gravitomagnetic approach.

For lack of something better, the electromagnetic analogues of (1.7) are often applied in cases, where $R \leq r_0$ (see, e.g., Michel and Li [10]). Note, however, that the results of (1.7) show the flaw of a singularity at $R = 0$. In this work the validity of the gravitomagnetic fields of (1.7) and the corresponding electromagnetic fields at short distances from the sphere will be investigated. Especially, the validity of the expressions for fields $\mathbf{B}_p$(sphere) and $\mathbf{B}_{eq}$(sphere) from (1.7) at distance $R = r_0$ will be considered. In the gravitomagnetic case the respective fields of (1.7) reduce to

$$\mathbf{B}_p(\text{sphere}) = -\tfrac{2}{5}\beta c^{-1} G^{\frac{1}{2}} m r_0^{-1} \boldsymbol{\omega}, \quad \mathbf{B}_{eq}(\text{sphere}) = \tfrac{1}{5}\beta c^{-1} G^{\frac{1}{2}} m r_0^{-1} \boldsymbol{\omega}. \tag{1.8}$$

Using series expansions, we will try to calculate more accurate values for the gravitomagnetic field $\mathbf{B}$(gm) inside and outside a sphere with a homogeneous mass density. Analogously, the corresponding electromagnetic induction fields $\mathbf{B}$(em) for a sphere with a homogeneous charge density have been deduced.

In general, the tiny effects due the gravitomagnetic field are difficult to observe. Two such effects have been investigated by space missions: the Gravity Probe B (GP-B) mission and the LAGEOS satellites. We will give a short introduction of both missions.

We will first consider the GP-B mission. Starting from the theory of general relativity, a derivation of the precession rate of a gyroscope moving in an orbit around a large massive rotating sphere has been given by Schiff [11, 12]. For a gyroscope in free fall, as in a satellite, the found precession rate consisted of two terms: the geodetic term and the frame-dragging term. We will only focus on the latter term. Owing to a formal analogy between electrodynamics and the general theory of relativity in the slow motion and weak field approximation, this contribution is often denoted as the gravitomagnetic term, whereas Weinberg [13, chs. 5 and 9] called it the spin-spin or "hyperfine" term. This contribution to the precession rate has previously been written in terms of the gravitomagnetic field $\mathbf{B}$ ([2, ch. 4], [3] and [5]). In ref. [5] both the classical and the gravitomagnetic derivation are given and discussed. Using the gravitomagnetic approach, one obtains

$$\boldsymbol{\Omega}(\text{gm}) = -2\beta^{-1} c^{-1} G^{\frac{1}{2}} \mathbf{B}, \tag{1.9}$$

where $\boldsymbol{\Omega}$(gm) is the precession rate of the angular momentum $\mathbf{S}$ around the direction of the field $\mathbf{B}$ ($\mathbf{S}$ is given by (1.3)). Combination of (1.5), (1.6) and (1.9) leads to the standard expression for the "hyperfine" precession rate $\boldsymbol{\Omega}$(gm)

$$\boldsymbol{\Omega}(\text{gm}) = c^{-2} G \left( \frac{3 \mathbf{S} \cdot \mathbf{R}}{R^5} \mathbf{R} - \frac{\mathbf{S}}{R^3} \right). \tag{1.10}$$

Note that this result does not depend on the constant $\beta$.

The prediction $\boldsymbol{\Omega}$(gm) from (1.10) can be tested by a gyroscope in an orbit moving around a rotating body with a large angular momentum $\mathbf{S}$. In order to do so, the Gravity Probe B spacecraft [14] was launched in a polar orbit around the Earth on April 20, 2004. Equipped with a set of four spherical gyroscopes and a telescope, it was designed to measure the precessions of the four gyroscopes with respect to a distant fixed star. Since the gravitomagnetic field $\mathbf{B}$ in (1.9) (see (1.6b)) has different values, at the pole and at the equator, for example, $\boldsymbol{\Omega}$(gm) of (1.10) has to be integrated over a revolution. The latter integration yields

$$\boldsymbol{\Omega}(\text{gm}) = \frac{G \mathbf{S}}{c^2 R^3} \left\{ \int_0^{2\pi} (3\cos^2\vartheta - 1) d\vartheta \bigg/ \int_0^{2\pi} d\vartheta \right\} = \frac{G \mathbf{S}}{2 c^2 R^3}, \tag{1.11}$$

where $\vartheta$ is the angle between the directions of $\mathbf{S}$ and $\mathbf{R}$. Calculation from (1.11) shows, that $\boldsymbol{\Omega}$(gm) equals 40.8 milliarc-seconds per year (mas.yr$^{-1}$).

As a second application of (1.9), a small perturbation of the orbit of a satellite around, e.g., the Earth can be considered. The orbiting satellite can be regarded as a gyroscope with angular momentum **S**$_{orbit}$, which is subjected to the gravitomagnetic field **B** (and the gravity field of the Earth). The precession of **S**$_{orbit}$ around field **B** is called the Lense-Thirring precession. For a polar orbit, however, the field **B** is no constant and **B** has to be integrated over the whole orbit. For an orbit with semi-major axis $a$ and orbital eccentricity $e$, the averaged value for the frequency rate, $\overline{\mathbf{\Omega}_{LT}}$, can be shown to be (see, e.g., ref. [15] for the original derivation)

$$\overline{\mathbf{\Omega}_{LT}} = \frac{2G\mathbf{S}}{c^2 a^3 (1-e^2)^{3/2}}. \tag{1.12}$$

Ciufolini *et al.* [16–19] tried to test the prediction of (1.12) by analysing the relativistic effects on the orbits of two artificial satellites: LAGEOS (laser geodynamics satellite) and LAGEOS 2. Recently, Ciufolini and Pavlis [18] reported a result of 99 ± 5 per cent of the value predicted by (1.12), but they allowed for a total error of ± 10 per cent to include unknown sources of error.

Analogously to the Lense-Thirring precession rate of an orbiting satellite, the precession rate of a circular torus with electrically neutral mass moving around a star could be tested. Following the same gravitomagnetic approach, the latter precession rate has previously been deduced [20, see eqs. (1.8) and (1.9) therein]. However, the predicted precession frequencies for pulsars like SAX J1808.4–3658, XTE J1807–294 and IGR J00291+5934 have not yet unambiguously be attributed to observed low frequency quasi-periodic oscillations (QPOs) (see ref. [20]). Perhaps, evidence for the existence of such low Lense-Thirring precession frequencies will be found in the future.

In section 2 the gravitomagnetic vector potential **A** of a torus with a homogeneous mass density, necessary for the subsequent calculation of the gravitomagnetic field, is deduced. In sections 3 and 4 the polar and equatorial gravitomagnetic fields, **B**$_p$(sphere) and **B**$_{eq}$(sphere), respectively, are calculated from **A**. An additional check of **B**$_{eq}$(sphere) is performed by application of Stokes' theorem. In section 5 the calculated fields are used in order to calculate the value of the gravitomagnetic precession rate for the gyroscopes in the Gravity Probe B satellite. Subsequently, the observed results from the GP-B mission will be compared with obtained predictions and the standard prediction from (1.11). Likewise, in section 6 observed data from the LAGEOS satellites will be compared with the value of the Lense-Thirring prediction from (1.12). In section 7 electromagnetic analogues of the gravitomagnetic field **B**(gm), **B**(em), are shortly discussed. In addition, the nature of **B**(gm) is considered more in detail. Finally, conclusions are drawn in section 8.

## 2. THE GRAVITOMAGNETIC VECTOR POTENTIAL A

Firstly, the gravitomagnetic vector potential **A** from (1.1) for a circular torus containing a total mass $dm = \rho dV$ (where $\rho$ is the homogeneous mass density of the sphere) will be considered. The derivation of **A** follows from the method given by Jackson [21]. As an example, a torus lying in an $x'$-$y'$ plane at distance $s$ from the origin $O'$ is chosen, as shown in figure 1. The $x'$-$y'$ plane is parallel to the $x$-$y$ plane through the centre $O$ of the sphere. A radius vector **R'** from $O'$ to a field point $F$ is fixed by the spherical coordinates $R'$, $\theta$ and $\varphi' = 0$. At field point $F$, the mass current $dm/dt = dm\,v$ ($v$ is the frequency of the mass current) in the torus generates the following azimuthal component of the vector potential **A** in the $y$ direction, i.e. $A_{\varphi'}(R',\theta)$

$$A_{\varphi'}(R',\theta) = \frac{-8\beta G^{1/2} dm\, v\, s}{c(s^2 + R'^2 + 2sR'\sin\theta)^{1/2}} \left\{ \frac{(1-\tfrac{1}{2}k^2)K(k) - E(k)}{k^2} \right\}, \tag{2.1}$$

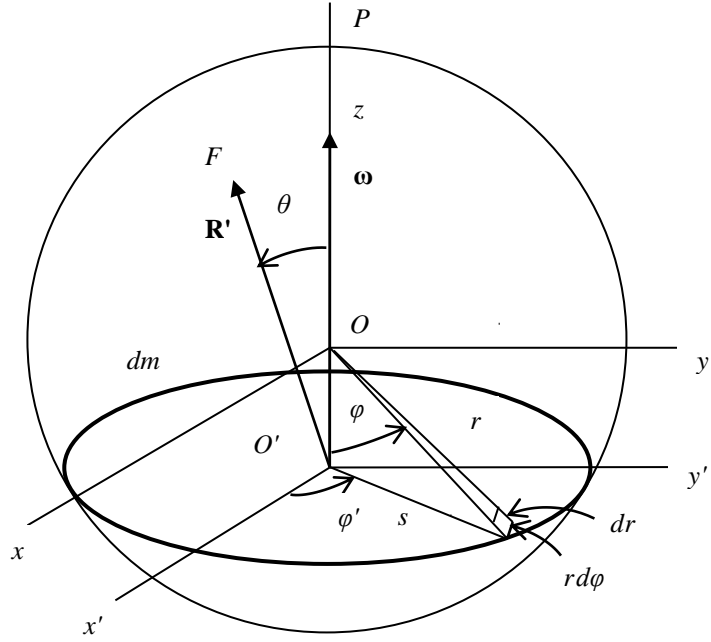

Figure 1. Spherical coordinates $R'$, $\theta$ and $\varphi' = 0$ of a field point $F$ relative to the origin $O'$. A point mass $\delta m$ in a torus is located at coordinates $s$ and $\varphi' = \varphi'$ in the $x'$-$y'$ plane. The total mass $dm$ of the torus is given by $\sum \delta m = dm = \rho \, dV = 2\pi \rho \, s \, r d\varphi \, dr$. The angular velocity vector of the sphere is denoted by $\boldsymbol{\omega}$ ($\omega = 2\pi \nu$ is the angular velocity of the sphere, where $\nu$ is its rotational frequency). The distance $OP = R$ denotes a position at or above the pole of the sphere.

where $K(k)$ and $E(k)$ are complete elliptic integrals of the first kind and second kind, respectively. See for the properties of these integrals, e.g., [22, § 2.57, § 8.11–§ 8.12]. The modulus $k$ of the elliptic integrals is given by

$$k^2 = \frac{4sR'\sin\theta}{s^2 + R'^2 + 2sR'\sin\theta}. \tag{2.2}$$

A number of limiting cases for $A_{\varphi'}(R', \theta)$ can be distinguished. When $\theta \approx 0$, $R' \gg s$, or $s \gg R'$, $k^2$ is small. Then, $A_{\varphi'}(R', \theta)$ in (2.1) reduces to relative simple expressions. For $k = 1$ $K(k)$ and $A_{\varphi'}(R', \theta)$ become infinite. We will now first treat the limiting case $\theta \approx 0$. In section 4 an example of $R' \gg s$, or $R' \to \infty$ with $F$ in the $x$-$y$ plane will be considered.

In the limiting case $\theta \approx 0$, the value of $k^2$ is small and the expression between parentheses in (2.1), $F(k)$, reduces to

$$F(k) \equiv \frac{(1 - \tfrac{1}{2}k^2)K(k) - E(k)}{k^2} \approx \frac{\pi k^2}{32}. \tag{2.3}$$

Combination of (2.1)–(2.3) then yields for $A_{\varphi'}(R', \theta)$ (compare with Jackson [21])

$$A_{\varphi'}(R', \theta) = \frac{-\beta G^{1/2} dm \, \omega \, s^2 R' \sin\theta}{2c(s^2 + R'^2 + 2sR'\sin\theta)^{3/2}}, \tag{2.4}$$

where the relation $\omega = 2\pi\nu$ has been inserted.

When values of $\theta \gg 0$ occur, the approximate value of $F(k)$ in (2.3) is no longer valid and a series expansion of $F(k)$ in terms of $k^2$ can be applied. In order to calculate $F(k)$, series expansions of the complete elliptic integrals $K(k)$ and $E(k)$, up to the terms in

$k^{14}$, are performed (see, e.g., ref. [22, § 8.11]). One obtains for $F(k)$

$$F(k) = \frac{\pi k^2}{32}\left(1+\frac{3}{4}k^2+\frac{75}{128}k^4+\frac{245}{512}k^6+\frac{6615}{2^{14}}k^8+\frac{22869}{2^{16}}k^{10}+\frac{1288287}{2^{22}}k^{12}+\ldots\right). \quad (2.5)$$

Instead of the parameters $R'$ and $s$ occurring in (2.1) and (2.2), the coordinates $R$, $r$ and $\varphi$ will be used in the evaluation of these equations below

$$R \equiv R'\sin\theta, \quad O'O = r\cos\varphi, \quad R'^2 = r^2\cos^2\varphi + R^2 \quad \text{and} \quad s = r\sin\varphi. \quad (2.6)$$

These relations between $R'$ and $R$ will be used for a field point $F$, lying in the equatorial plane that coincides with the $x$-$y$ plane in figure 1. Moreover, the following relation will be introduced into (2.1)

$$dm = 2\pi\rho\sin\varphi\, d\varphi\, r^2 dr. \quad (2.7)$$

By combining (2.2), (2.3a) and (2.5)–(2.7) with (2.1), $A_{\varphi'}(R',\theta)$ transforms into

$$A_\varphi(r,\varphi) = \frac{-\pi\beta G^{\frac{1}{2}}\rho\omega R r^4 dr \sin^3\varphi\, d\varphi}{c(r^2+R^2+2rR\sin\varphi)^{\frac{3}{2}}}f(k), \quad (2.8)$$

where $f(k)$ is defined by the relation $f(k) \equiv (32/\pi k^2)F(k)$; $F(k)$ is given by (2.5). Moreover, introduction of (2.6) into $k^2$ of (2.2) transforms $k^2$ into

$$k^2 = \frac{4rR\sin\varphi}{r^2+R^2+2rR\sin\varphi}. \quad (2.9)$$

Note that $k$ obtains unity value, when $\varphi = \frac{1}{2}\pi$ and $r = R$. The quantity $f(k)$ then becomes infinite.

In sections 3 and 4 we will calculate the polar and equatorial gravitomagnetic fields from a sphere, $\mathbf{B}_p$(sphere) and $\mathbf{B}_{eq}$(sphere), from equations (2.4) and (2.8), respectively.

## 3. THE POLAR GRAVITOMAGNETIC FIELD OF A SPHERE

In order to deduce $\mathbf{B}_p$(sphere), we first calculate the radial component $B_{R'}$ of the gravitomagnetic induction field $\mathbf{B}$ from the relation

$$B_{R'} = \frac{1}{R'\sin\theta}\frac{\partial}{\partial\theta}\{\sin\theta A_{\varphi'}(R',\theta)\}, \quad (3.1)$$

where $R'$, $\theta$ and $\varphi'$ have been defined in section 2 (see also figure 1). Application of (3.1) to the approximated azimuthal component of the vector potential of (2.4) from a torus with total mass $dm$ yields (compare with Jackson [21])

$$B_{R'}(R',\theta) \approx \frac{-\beta G^{\frac{1}{2}}dm\,\omega\cos\theta\, s^2(2s^2+2R'^2+sR'\sin\theta)}{2c(s^2+R'^2+2sR'\sin\theta)^{\frac{5}{2}}}. \quad (3.2)$$

Next, we will consider the exact limiting case $\theta = 0$ by choosing a field point $P$ at or above the pole of the sphere, so that $OP = R$ ($R \geq r_0$, see also figure 1). Then, $R'$ and $s$ can be replaced by

$$R' = r\cos\varphi + R \quad \text{and} \quad s = r\sin\varphi. \tag{3.3}$$

Combination of (2.7), (3.2) and (3.3) then transforms $B_{R'}(R', \theta)$ from (3.2) into

$$B_R(r,\varphi) = \frac{-2\pi\beta G^{\frac{1}{2}}\rho\omega\sin^3\varphi\,d\varphi\,r^4 dr}{c(r^2 + R^2 + 2rR\cos\varphi)^{\frac{3}{2}}}. \tag{3.4}$$

The gravitomagnetic induction field at field point $P$ from a thin shell with thickness $dr$, $B_p(r, \text{shell})$, can now be calculated from (3.4) by integration of $\varphi$ from 0 to $\pi$

$$B_p(r,\text{shell}) = \frac{-2\pi\beta G^{\frac{1}{2}}\rho\omega r^4 dr}{c}\int_0^\pi \frac{\sin^3\varphi\,d\varphi}{(r^2 + R^2 + 2rR\cos\varphi)^{\frac{3}{2}}} = \frac{-8\pi\beta G^{\frac{1}{2}}\rho\omega r^4 dr}{3cR^3}. \tag{3.5}$$

In the integration of $\varphi$ the quantity $R$ has been considered as a constant parameter. It appears that the result of the integral does not depend on $r$. Integration of $B_p(r, \text{shell})$ from (3.5) over the interval $r = 0$ to $r = r_0$ ($r_0$ is the radius of the sphere) then yields the gravitomagnetic induction field at the poles of the sphere, $B_p(\text{sphere})$

$$B_p(\text{sphere}) = \frac{-8\pi\beta G^{\frac{1}{2}}\rho\omega}{3cR^3}\int_0^{r_0} r^4 dr = \frac{-8\pi\beta G^{\frac{1}{2}}\rho\omega r_0^5}{15cR^3}. \tag{3.6}$$

For a sphere with homogeneous mass density $\rho$ the total mass $m$ of the sphere is given by

$$m = \tfrac{4}{3}\pi\rho r_0^3. \tag{3.7}$$

Combination of (3.6) and (3.7) then leads to the following accurate result for $\mathbf{B}_p(\text{sphere})$

$$\mathbf{B}_p(\text{sphere}) = -\tfrac{2}{5}\beta c^{-1} G^{\frac{1}{2}} m r_0^2 R^{-3} \boldsymbol{\omega}. \tag{3.8}$$

Thus, the gravitomagnetic field $\mathbf{B}_p(\text{sphere})$ of (3.8) appears to coincide with the field from the ideal dipole approximation of (1.7a), deduced for the limiting case $R > r_0$. Moreover, relation (3.8) retains its validity up to distance $R = r_0$. In the latter case the field of (3.8) reduces to the expression of (1.8a).

## 4. THE GRAVITOMAGNETIC FIELD IN THE EQUATORIAL PLANE

In order to calculate the equatorial gravitomagnetic field of a sphere, $\mathbf{B}_{eq}(\text{sphere})$, we will use the series expansion of $A_\varphi(r, \varphi)$ from (2.8), deduced for a torus, and that of $k^2$ from (2.9). In these series expansions the following dimensionless parameters $x$ and $y$ are introduced, defined by

$$x \equiv \frac{2rR(\sin\varphi - 1)}{(r+R)^2} = y(\sin\varphi - 1), \quad y \equiv \frac{2rR}{(r+R)^2}. \tag{4.1}$$

Note that both $x$ and $y$ have values smaller than unity, so that series expansions in $x$ or $y$ may converge. As an example, the factor $1/(r^2 + R^2 + 2rR\sin\varphi)^{3/2}$ in (2.8) has been expanded up to sixth order terms in $x$. One obtains

$$\frac{1}{(r+R)^3(1+x)^{3/2}} = \frac{1}{(r+R)^3}\left(1 - \frac{3}{2}x + \frac{15}{8}x^2 - \frac{35}{16}x^3 + \frac{315}{128}x^4 - \frac{693}{256}x^5 + \frac{3003}{1024}x^6 - \ldots\right). \quad (4.2)$$

Likewise, the terms in $k^2$, $k^4$, and so on, in the function $f(k)$ from (2.8) have also been expanded up to sixth order terms in $x$. After introduction of all these series expansions into (2.8), integration of $\varphi$ from 0 to $\pi$ follows. The following result for $A_\varphi(r, \text{shell})$ is then obtained after a lengthy but straightforward calculation

$$A_\varphi(r, \text{shell}) = \frac{-\pi\beta G^{1/2}\rho\omega R r^4 dr}{c} \int_0^\pi \frac{\sin^3\varphi f(k) d\varphi}{(r^2+R^2+2rR\sin\varphi)^{3/2}} = \frac{-4\pi\beta G^{1/2}\rho\omega R r^4 g(y) dr}{3c(r+R)^3}, \quad (4.3)$$

where the quantity $g(y)$ is given by

$$g(y) = \left(1 + \frac{3}{2}y + \frac{9}{4}y^2 + \frac{7}{2}y^3 + \frac{45}{8}y^4 + \frac{297}{32}y^5 + \frac{1001}{64}y^6 + \ldots\right). \quad (4.4)$$

Like the terms in $x$ in (4.2), for example, the terms in $y$ in (4.4) have been calculated up to sixth order. Expression $A_\varphi(r, \text{shell})$ from (4.3) can further be evaluated by integration of $r$ from 0 to $r_0$. After substitution of (3.7) and a lengthy calculation then follows for the azimuthal component of the vector potential **A** of the sphere, $A_\varphi(\text{sphere})$

$$A_\varphi(\text{sphere}) = \frac{-\beta G^{1/2}m\omega R}{cr_0^3}\int_0^{r_0}\frac{r^4 g(y)dr}{(r+R)^3} = \frac{-\beta G^{1/2}m\omega R}{cr_0}\left[\left\{\frac{-\frac{5}{2}r_0^2 - 2r_0 R}{(r_0+R)^2} + 3 - 6\frac{R}{r_0}\right.\right.$$

$$+ 6\frac{R^2}{r_0^2}\ln\left(\frac{r_0+R}{R}\right)\right\} + \left\{\frac{-12r_0^3 R - \frac{115}{4}r_0^2 R^2 - 25r_0 R^3 - \frac{15}{2}R^4}{(r_0+R)^4} + 15\frac{R}{r_0} - 15\frac{R^2}{r_0^2}\ln\left(\frac{r_0+R}{R}\right)\right\}$$

$$+ \left\{\frac{-\frac{441}{20}r_0^5 R^2 - \frac{783}{10}r_0^4 R^3 - \frac{513}{4}r_0^3 R^4 - 111r_0^2 R^5 - \frac{99}{2}r_0 R^6 - 9R^7}{r_0(r_0+R)^6} + 9\frac{R^2}{r_0^2}\ln\left(\frac{r_0+R}{R}\right)\right\} \quad (4.5)$$

$$+ \frac{7}{2}\frac{R^2}{r_0^2}\left\{\frac{r_0^8}{(r_0+R)^8}\right\} + \frac{R^2}{r_0^2}\left\{\frac{r_0^{10} + 10r_0^9 R}{(r_0+R)^{10}}\right\} + \frac{9}{20}\frac{R^2}{r_0^2}\left\{\frac{r_0^{12} + 12r_0^{11}R + 66r_0^{10}R^2}{(r_0+R)^{12}}\right\}$$

$$\left.+ \frac{1}{4}\frac{R^2}{r_0^2}\left\{\frac{r_0^{14} + 14r_0^{13}R + 91r_0^{12}R^2 + 364r_0^{11}R^3}{(r_0+R)^{14}}\right\}\right].$$

It is noted that the seven terms between parentheses on the right hand side of (4.5) correspond to the unity term, the terms in $y$, $y^2$, up to $y^6$, in the series expansion on the r. h. s. of (4.4), respectively. In addition, it is noticed that the last four terms between parentheses, display an increasing number of terms in the nominator that also occur in the denominator. For example, $(r_0+R)^{14}$ can be written as

$$(r_0+R)^{14} = r_0^{14} + 14r_0^{13}R + 91r_0^{12}R^2 + 364r_0^{11}R^3 + 1001r_0^{10}R^4 + \ldots + R^{14}. \quad (4.6)$$

Comparison of (4.5) and (4.6) shows that only the first four terms out of fifteen in the series expansion on the right hand side of (4.6) occur in the nominator of the corresponding series of (4.5).

Two applications of (4.5) will separately be investigated: expressions of $A_\varphi(\text{sphere})$ inside and outside the sphere, respectively. Firstly, for $R \leq r_0$ series expansion of the right hand side of (4.5) up to terms $(R/r_0)^7$ yields

$$A_\varphi(\text{sphere}) = \frac{-\beta G^{1/2} m\omega}{c}\left(\frac{1}{2}\frac{R}{r_0} - \frac{11}{10}\frac{R^3}{r_0^3}\right). \quad (4.7)$$

Apart from the terms in $R/r_0$ and $(R/r_0)^3$, all terms of the truncated series expansion, like terms in $(R/r_0)^2$, $(R/r_0)^4$, $(R/r_0)^5$, $(R/r_0)^6$ and $(R/r_0)^7$ reduce to zero value. Moreover, the coefficient 11/5 of the surviving term in $(R/r_0)^3$ will certainly decrease, when higher order terms than $y^6$ are added to the calculation of (4.5). For small values of $R$ a linear relationship between $A_\varphi(\text{sphere})$ and $R$ can be calculated from (4.7)

$$A_\varphi(\text{sphere}) = \frac{-\beta G^{1/2} m\omega R}{2cr_0}. \quad (4.8)$$

In addition, it follows from this relation that $A_\varphi(\text{sphere})$ reduces to zero value in the limiting case $R \to 0$.

Secondly, for $R \geq r$ series expansion of the r. h. s. of $A_\varphi(\text{sphere})$ in (4.5) yields the following result for the limiting case $R \to \infty$

$$A_\varphi(\text{sphere}, R \to \infty) = \frac{-\beta G^{1/2} m\omega r_0^2}{5cR^2}. \quad (4.9)$$

Apart from the term in $(r_0/R)^2$ in (4.9) the series expansion of $A_\varphi(\text{sphere})$ from (4.5) also produces terms in $(R/r_0)^2$, $R/r_0$, integers, $r_0/R$, $(r_0/R)^2$, $(r_0/R)^3$, $(r_0/R)^4$, $(r_0/R)^6$, $(r_0/R)^7$ and $(r_0/R)^8$, but their sums all cancel (higher order terms have not been calculated). Only the term in $(r_0/R)^2$ survives. The results of (4.8) and (4.9) lead to important limiting cases for the gravitomagnetic field. We will calculate these fields below.

The $\theta$-component of the gravitomagnetic induction field **B**, $B_\theta$, can be deduced from the relation

$$B_\theta = -\frac{1}{R}\frac{\partial}{\partial R}\{RA_\varphi(\text{sphere})\}. \quad (4.10)$$

Introduction of (4.5) into (4.10), followed by evaluation, then yields for the equatorial gravitomagnetic induction field, $\mathbf{B}_{eq}(\text{sphere})$, at distance $R$ from the centre of the sphere

$$\begin{aligned}
\mathbf{B}_{eq}(\text{sphere}) = \frac{-\beta G^{1/2} m\boldsymbol{\omega}}{cr_0}\Bigg\langle &\left\{\frac{+r_0^4 - 6r_0^3 R - 44r_0^2 R^2 - 60r_0 R^3 - 24R^4}{r_0(r_0+R)^3} + 24\frac{R^2}{r_0^2}\ln\left(\frac{r_0+R}{R}\right)\right\} \\
&+\left\{\frac{+9r_0^5 R + 137r_0^4 R^2 + 385r_0^3 R^3 + 470r_0^2 R^4 + 270r_0 R^5 + 60R^6}{r_0(r_0+R)^5} - 60\frac{R^2}{r_0^2}\ln\left(\frac{r_0+R}{R}\right)\right\} \\
&+\left\{\frac{9R^2}{r_0^2}\frac{\left(-\frac{54}{5}r_0^7 - \frac{223}{5}r_0^6 R - \frac{459}{5}r_0^5 R^2 - \frac{319}{3}r_0^4 R^3 - \frac{214}{3}r_0^3 R^4 - 26r_0^2 R^5 - 4r_0 R^6\right)}{(r_0+R)^7} + 36\frac{R^2}{r_0^2}\ln\left(\frac{r_0+R}{R}\right)\right\} \\
&+\left\{\frac{14R^2}{r_0^2}\frac{(r_0^9 - r_0^8 R)}{(r_0+R)^9}\right\} + \left\{\frac{4R^2}{r_0^2}\frac{(r_0^{11}+11r_0^{10}R - \frac{25}{2}r_0^9 R^2)}{(r_0+R)^{11}}\right\} + \left\{\frac{9}{5}\frac{R^2}{r_0^2}\frac{(r_0^{13}+13r_0^{12}R+78r_0^{11}R^2-99r_0^{10}R^3)}{(r_0+R)^{13}}\right\} \\
&+\left\{\frac{R^2}{r_0^2}\frac{(r_0^{15}+15r_0^{14}R+105r_0^{13}R^2+455r_0^{12}R^3-637r_0^{11}R^4)}{(r_0+R)^{15}}\right\}\Bigg\rangle.
\end{aligned} \quad (4.11)$$

This complicated expression can be regarded as a main result of this work. It is noticed that the logarithmic terms in (4.11) cancel.

Two limiting cases for (4.11) will now be distinguished. Firstly, for $R \leq r_0$ series expansion of $\mathbf{B}_{eq}$(sphere) from (4.11) up to terms in $(R/r_0)^6$ yields

$$\mathbf{B}_{eq}(\text{sphere}) = \frac{-\beta G^{\frac{1}{2}} m \omega}{cr_0}\left(1 - \frac{22}{5}\frac{R^2}{r_0^2}\right). \tag{4.12}$$

It appears that all terms like terms in $R/r_0$ $(R/r_0)^3$, $(R/r_0)^4$, up to terms in $(R/r_0)^6$ reduce to zero value. When higher order terms than $y^6$ are added to (4.4), a calculation leading to the analogue of (4.12), may yield the limit value 6/5 instead of the factor 22/5. In that case the value of $\mathbf{B}_{eq}$(sphere) changes into

$$\mathbf{B}_{eq}(\text{sphere}) = \frac{-\beta G^{\frac{1}{2}} m \omega}{cr_0}\left(1 - \frac{6}{5}\frac{R^2}{r_0^2}\right). \tag{4.13}$$

For $R = r_0$ the field $\mathbf{B}_{eq}$(sphere) reduces to value of the ideal gravitomagnetic dipole of (1.7b). In the limiting case of $R \to 0$ the series expansion of $\mathbf{B}_{eq}$(sphere) of both (4.12) and (4.13) simplify to $\mathbf{B}_c$(sphere), the gravitomagnetic field at the centre of the sphere

$$\mathbf{B}_c(\text{sphere}) = \frac{-\beta G^{\frac{1}{2}} m \omega}{cr_0}. \tag{4.14}$$

The latter result can more easily be found by a direct calculation from (4.8) and (4.10). Relation (4.14) has earlier been given by Biemond [2]. Contrary to the ideal dipole model discussed in section 1, no singularity is obtained for $\mathbf{B}_c$(sphere) (see comment to (1.7)).

Secondly, another limiting case occurs for $R \geq r_0$. The series expansion of (4.11) then yields for $\mathbf{B}_{eq}$(sphere)

$$\mathbf{B}_{eq}(\text{sphere}) = \frac{\beta G^{\frac{1}{2}} m r_0^2 \omega}{5cR^3}\left\{1 + \text{terms in }\left(\frac{r_0}{R}\right)^n\right\}. \tag{4.15}$$

The series expansion on the r. h. s. of (4.11) may contain non-zero terms in $(r_0/R)^n$ with $n$ ranging from $n = -4$ up to $+\infty$. Although for $n = -4, -3, -2, -1, +1, +2, +3, +4, +5$ and $+6$ appear to be zero, higher order terms may give a non-zero contribution. When all terms in $(R/r_0)^n$ for $n \geq +7$ would be zero, only the first term on the r. h. s. of in (4.15) would survive. In that case the relation of (4.15) would coincide with the result of (1.7b) for the ideal dipole model. In any case, for $R \gg r_0$, the first term of (4.15) reduces to the result of (1.7b). By combining (4.9) and (4.10), this limiting value of (4.15) follows more directly.

Furthermore, calculations of the sum $S$ of the seven terms between parentheses on the r.h.s. of (4.11) (term $1 \equiv T1$, and so on) show that $S$ yields zero value for $R = 1.0758 r_0$. When more higher order terms are included in the truncated series expansion of (4.11), the value $S = 0$ shifts to lower values of $R$. In order to get an impression of this change, we calculate the value of $R$, where the subsequent terms between parentheses on the r. h. s. of (4.11) reduce to zero value. One obtains

$$\begin{aligned}&\text{term} 1: R = 1.5865 \, r_0, \text{ term} 2: R = 1.1923 \, r_0, \text{ term} 3: R = 1.0632 \, r_0, \text{ term} 4: R = r_0, \\ &\text{term} 5: R = 0.9631 \, r_0, \text{ term} 6: R = 0.9392 \, r_0, \text{ term} 7: R = 0.9226 \, r_0.\end{aligned} \tag{4.16}$$

These results suggest that for a more extended series expansion of (4.11) $\mathbf{B}_{eq}$(sphere) reduces to zero value for $R < r_0$. For the hypothetical expression (4.13) $\mathbf{B}_{eq}$(sphere) reduces to zero value for $R = (5/6)^{\frac{1}{2}} r_0 = 0.9129 \, r_0$.

Table 1. Calculated contributions to **B**$_{eq}$(sphere), in units of $-\beta c^{-1}G^{\frac{1}{2}}mr_0^{-1}\omega$, for different values of $R$ from the respective terms between parentheses on the r. h. s. of (4.11). In addition, the sum of the seven contributions to **B**$_{eq}$(sphere), $S$, in units of $-\beta c^{-1}G^{\frac{1}{2}}mr_0^{-1}\omega$, is given. Moreover, the values of **B**$_{eq}$(sphere) from (4.13), $E$, (educated guess) and (1.7b), (ideal dipole model) in units of $-\beta c^{-1}G^{\frac{1}{2}}mr_0^{-1}\omega$, are also calculated for values of $0 \leq R \leq r_0$ and $R \geq r_0$, respectively. The ratio $S/D$ in percent is added, too. See also text.

| Term number | $R = 0$ | $R = 0.1\,r_0$ | $R = 0.2\,r_0$ | $R = 0.3\,r_0$ | $R = 0.4\,r_0$ | $R = 0.5\,r_0$ |
|---|---|---|---|---|---|---|
| 1 | +1 | +0.4985602 | +0.2858298 | +0.1748438 | +0.1114844 | +0.0731552 |
| 2 | 0 | +0.2407060 | +0.2017063 | +0.1432946 | +0.0968765 | +0.0640255 |
| 3 | 0 | +0.1108213 | +0.1321591 | +0.1093745 | +0.0793897 | +0.0537657 |
| 4 | 0 | +0.0534363 | +0.0868254 | +0.0831722 | +0.0650499 | +0.0455215 |
| 5 | 0 | +0.0276890 | +0.0581420 | +0.0637778 | +0.0537342 | +0.0390184 |
| 6 | 0 | +0.0155428 | +0.0398919 | +0.0494598 | +0.0447903 | +0.0338160 |
| 7 | 0 | +0.0094352 | +0.0280932 | +0.0388149 | +0.0376558 | +0.0295805 |
| $S$ | +1 | +0.9561908 | +0.8326477 | +0.6627376 | +0.4889809 | +0.3388830 |
| $E$ | +1 | +0.988 | +0.952 | +0.892 | +0.808 | +0.700 |

| Term number | $R = 0.6\,r_0$ | $R = 0.7\,r_0$ | $R = 0.8\,r_0$ | $R = 0.9\,r_0$ | $R = r_0$ | $R = 1.1\,r_0$ |
|---|---|---|---|---|---|---|
| 1 | +0.0489741 | +0.0332201 | +0.0226920 | +0.0155112 | +0.0105323 | +0.0070346 |
| 2 | +0.0415967 | +0.0264679 | +0.0163027 | +0.0094834 | +0.0049192 | +0.0018792 |
| 3 | +0.0345870 | +0.0210692 | +0.0118577 | +0.0057277 | +0.0017360 | −0.0007977 |
| 4 | +0.0293367 | +0.0173542 | +0.0090341 | +0.0035142 | 0 | −0.0021327 |
| 5 | +0.0253749 | +0.0147264 | +0.0071699 | +0.0021555 | −0.0009766 | −0.0027981 |
| 6 | +0.0222964 | +0.0127902 | +0.0058820 | +0.0012859 | −0.0015381 | −0.0031145 |
| 7 | +0.0198356 | +0.0113103 | +0.0049560 | +0.0007101 | −0.0018616 | −0.0032412 |
| $S$ | +0.2220014 | +0.1369383 | +0.0778944 | +0.0383881 | +0.01281128 | −0.0031704 |
| $D$ |  |  |  |  | −0.2 | −0.1502630 |
| $E$ | +0.568 | +0.412 | +0.232 | +0.028 | −0.2 |  |
| $S/D$ (%) |  |  |  |  |  | 2.110 |

| Term number | $R = 1.2\,r_0$ | $R = 1.3\,r_0$ | $R = 1.4\,r_0$ | $R = 1.5\,r_0$ | $R = 1.6\,r_0$ | $R = 1.7\,r_0$ |
|---|---|---|---|---|---|---|
| 1 | +0.0045522 | +0.0027772 | +0.0015019 | +0.0005837 | −0.0000766 | −0.0005492 |
| 2 | −0.0001264 | −0.0014275 | −0.0022476 | −0.0027392 | −0.0030067 | −0.0031222 |
| 3 | −0.0023488 | −0.0032441 | −0.0037063 | −0.0038861 | −0.0038854 | −0.0037726 |
| 4 | −0.0033398 | −0.0039408 | −0.0041547 | −0.0041288 | −0.0039606 | −0.0037141 |
| 5 | −0.0037459 | −0.0041327 | −0.0041733 | −0.0040108 | −0.0037385 | −0.0034156 |
| 6 | −0.0038633 | −0.0040909 | −0.0040081 | −0.0037541 | −0.0034177 | −0.0030533 |
| 7 | −0.0038340 | −0.0039408 | −0.0037688 | −0.0034534 | −0.0030788 | −0.0026950 |
| $S$ | −0.0127060 | −0.0179997 | −0.0205570 | −0.0213887 | −0.0211643 | −0.0203219 |
| $D$ | −0.1157407 | −0.0910332 | −0.0728863 | −0.0592593 | −0.0488281 | −0.0407083 |
| $S/D$ (%) | 10.98 | 19.77 | 28.20 | 36.09 | 43.34 | 49.92 |

| Term number | $R = 1.8\,r_0$ | $R = 1.9\,r_0$ | $R = 2.0\,r_0$ | $R = 2.5\,r_0$ | $R = 3.0\,r_0$ | $R = 10\,r_0$ |
|---|---|---|---|---|---|---|
| 1 | −0.0008840 | −0.0011174 | −0.0012755 | −0.0014677 | −0.0012974 | −0.00012144 |
| 2 | −0.0031356 | −0.0030815 | −0.0029840 | −0.0022806 | −0.0016394 | −0.00005344 |
| 3 | −0.0035935 | −0.0033786 | −0.0031479 | −0.0020704 | −0.0013235 | −0.00001751 |
| 4 | −0.0034304 | −0.0031354 | −0.0028451 | −0.0016653 | −0.0009613 | −0.00000534 |
| 5 | −0.0030785 | −0.0027488 | −0.0024387 | −0.0012850 | −0.0006738 | −0.00000160 |
| 6 | −0.0026930 | −0.0023543 | −0.0020458 | −0.0009758 | −0.0004661 | −0.00000047 |
| 7 | −0.0023296 | −0.0019964 | −0.0017008 | −0.0007367 | −0.0003212 | −0.00000014 |
| $S$ | −0.0191445 | −0.0178124 | −0.0164377 | −0.0104815 | −0.0066827 | −0.00019994 |
| $D$ | −0.0342936 | −0.0291588 | −0.025 | −0.0128 | −0.0074074 | −0.0002 |
| $S/D$ (%) | 55.83 | 61.09 | 65.75 | 81.89 | 90.22 | 99.97 |

In table 1 the separate values of the seven terms between parentheses on the r. h. s. of (4.11) for increasing values of $R$ have been given in units of $-\beta c^{-1}G^{\frac{1}{2}}mr_0^{-1}\omega$ for the interval $0 \leq R \leq 10r_0$. Furthermore, the sum $S$ of the seven terms has been added. The values of the hypothetical field $\mathbf{B}_{eq}$(sphere) from (4.13), denoted by $E$ (educated guess), for values of $R$ in the interval $0 \leq R \leq r_0$, have also been given in the same units. Finally, the value of $\mathbf{B}_{eq}$(sphere) from the dipole model of (1.7b), $D$, has been given again in the same units for values of $R \geq r_0$. The absolute value of the ratio $S/D$ in percent is also added to table 1. It appears that values of $D$ are always more negative than those of $S$. For values of $R \geq 10r_0$ both values coincide within 0.3%.

In figure 2 the first term (= term 1 = *T1*) between parentheses on the r. h. s. of $\mathbf{B}_{eq}$(sphere) of (4.11) and the $S$ have been plotted against increasing values of $R$. The former curve accurately displays the behavior of $\mathbf{B}_{eq}$(sphere) in the limiting cases $R \to 0$ and $R \to \infty$. The $S$ versus $R$ curve illustrates the more accurate overall behaviour of $\mathbf{B}_{eq}$(sphere), but is still a rough approximation. The result for $S$ would improve when higher order than $y^6$ would be incorporated into (4.4), but their calculation is cumbersome. The $S$ versus $R$ curve can be compared with $E$ versus $R$ curve from (4.13) for the interval $0 \leq R \leq r_0$ and with $D$ versus $R$ curve from (1.7b) for the interval $r_0 \leq R < 10r_0$ (or larger). The area between the $S$ one side and the $E$ and $D$ curves on the other reflects our uncertainty.

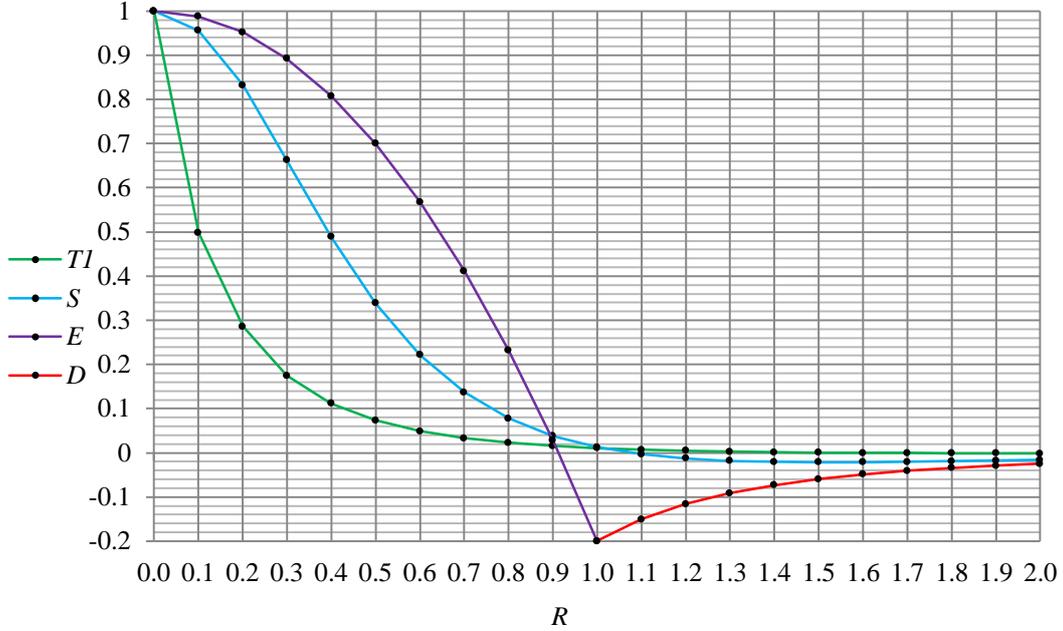

Figure 2. Values of *T1* and $S$ of $\mathbf{B}_{eq}$(sphere) from (4.11), expressed in units of $-\beta c^{-1}G^{\frac{1}{2}}mr_0^{-1}\omega$, have been plotted against increasing values of $R$, in units of $r_0$. Moreover, the values of $E$, the hypothetical expression of $\mathbf{B}_{eq}$(sphere) from (4.13), and the values $D$ from the dipole model of (1.7b), all expressed in units of $-\beta c^{-1}G^{\frac{1}{2}}mr_0^{-1}\omega$, have been plotted against $R$. See also text.

### 4.2 CALCULATION OF THE EQUATORIAL GRAVITOMAGNETIC FIELD BY APPLICATION OF STOKES' THEOREM

As an alternative method to calculate the equatorial gravitomagnetic field, one may use Stokes' theorem. For the surface denoted by *ABCDEFA* in figure 3 the r. h. s. of (1.1a) is zero, since $\rho \mathbf{v} = 0$. Application of Stokes' theorem to (1.1a) then yields

$$\oiint (\nabla \times \mathbf{B}) \cdot d\mathbf{S} = \oint \mathbf{B} \cdot d\mathbf{s} = \int_A^B \mathbf{B} \cdot d\mathbf{s} + \int_B^C \mathbf{B} \cdot d\mathbf{s} + \int_C^D \mathbf{B} \cdot d\mathbf{s} + \int_D^E \mathbf{B} \cdot d\mathbf{s} + \int_E^F \mathbf{B} \cdot d\mathbf{s} + \int_F^A \mathbf{B} \cdot d\mathbf{s} = 0. \quad (4.17)$$

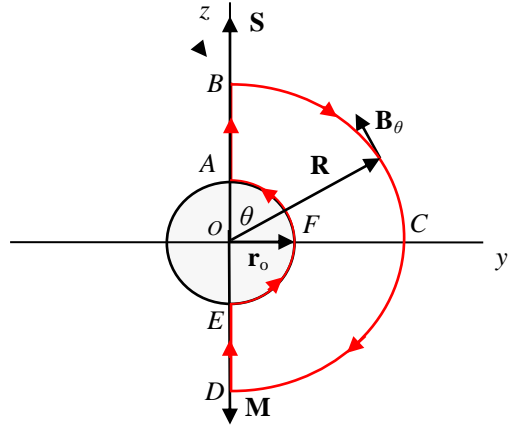

Figure 3. Application of Stokes' theorem to the surface *ABCDEFA* in the *y-z* plane through the centre of the sphere *O*. **S** denotes the angular momentum of the sphere and **M** is its gravitomagnetic moment. The angle between **S** and the position vector **R** is denoted by $\theta$. $B_\theta$ is the $\theta$-component of the gravitomagnetic field **B** along the semi-circle *BCD*.

Utilizing the calculated value for the gravitomagnetic field at the poles of the sphere, $\mathbf{B}_p$(sphere) of (3.8), relation (4.17) can be evaluated. By combining (1.3), (1.5) and (3.8), one obtains $\mathbf{B}_p$(sphere) $= 2\,R^{-3}\,\mathbf{M}$ for a homogeneous mass density (i.e., $f_s = 1$). Subsequent integration over the interval *AB* (and interval *DE*) then results into

$$\int_A^B \mathbf{B}\cdot d\mathbf{s} = \int_A^B \mathbf{B}_p(\text{sphere})\cdot d\mathbf{s} = \int_{r_0}^R 2R^{-3}\mathbf{M}\cdot d\mathbf{R} = \frac{M}{R^2} - \frac{M}{r_0^2} = \int_D^E \mathbf{B}\cdot d\mathbf{s}. \qquad (4.18)$$

When the radius *R* of the semi-circle *BCD* is very large, the gravitomagnetic field **B** along the semi-circle is given by the expression (1.6) for the ideal gravitomagnetic dipole located at the centre of the sphere. The component of **B** along the semi-circle *BCD*, is then given $B_\theta = -R^{-3}M\sin\theta$. Integration of **B** along the semi-circle *BCD* then yields

$$\int_B^C \mathbf{B}\cdot d\mathbf{s} + \int_C^D \mathbf{B}\cdot d\mathbf{s} = \int_0^\pi -MR^{-3}\sin\theta\, R\, d\theta = -\frac{2M}{R^2}. \qquad (4.19)$$

Combination of (4.17), (4.18) and (4.19) then yields the following result for the component of **B** along the semi-circle *EFA*

$$\int_E^F \mathbf{B}\cdot d\mathbf{s} + \int_F^A \mathbf{B}\cdot d\mathbf{s} = \frac{2M}{r_0^2}. \qquad (4.20)$$

As an example, the result of (4.17) can be obtained for the following component of the gravitomagnetic field $B_\theta$ along the semi-circle *EFA*

$$B_\theta = -M r_0^{-3} R \sin\theta. \qquad (4.21)$$

This expression for $B_\theta$ agrees with relation (1.6), representing the field of the ideal gravitomanetic dipole located at the centre of the sphere. Although the solution for $B_\theta$ is compatible with condition (4.17), a different result from (4.11) for $B_\theta$ might follow when a more complete series expansion would be calculated.

## 5. GRAVITY PROBE B

In the derivation of the precession rate $\mathbf{\Omega}$(gm) for a gyroscope of (1.11) the validity of a point dipole model is usually assumed. For example, the latter assumption has implicitly been made in the derivations of (1.10) given by Schiff [11, 12] and Weinberg [13], as well as in the gravitomagnetic approach followed by Biemond [5]. The last author compared the classical and gravitomagnetic derivations of (1.10) to each other. Whereas Weinberg [13] introduced a dimensionless vector potential $\zeta$ defined in his equation (9.1.61), Biemond [5] used the vector potential $\mathbf{A}$ defined in section 1 of this work. The vector potentials $\zeta$ and $\mathbf{A}$ are related by [5, p.4]

$$\zeta = 4\beta^{-1}c^{-2}G^{1/2}\mathbf{A}. \tag{5.1}$$

In this work the gravitomagnetic field $\mathbf{B}$ acting on the gyroscope has been calculated from the vector potential $\mathbf{A}$ (see eqs. (2.4) and (2.8)).

From the accurate calculation of the polar gravitomagnetic field $\mathbf{B}_p$(sphere) of (3.8), presented in section 3, follows that this result coincides with the field from the ideal dipole approximation (1.7a). The equatorial gravitomagnetic field $\mathbf{B}_{eq}$(sphere) from the truncated series expansion (4.11), however, deviates from the dipole field of (1.7b).

In order to compare observations with predictions from (1.11) we first calculate Earth's angular momentum $S$ from (1.3). Introduction of a homogeneity factor $f_s = 0.82675$ [23], a mass $m = 5.972 \times 10^{27}$ g, a radius $r_0 = 6,378$ km and an angular velocity $\omega = 7.292 \times 10^{-5}$ rad.s$^{-1}$ into (1.3), yields a value $S = 5.858 \times 10^{40}$ g.cm$^2$.s$^{-1}$.

In 2004 Gravity Probe B was launched in a nearly circular polar orbit with a semi-major axis of 7,027 km, corresponding to an orbit of about 649 km. Since the result for the precession rate $\mathbf{\Omega}$(gm) is only slightly affected by the small eccentricity $e = 0.0014$ (see data from 2005 in ref. [14]), we will neglect its influence on our calculation. Introduction of $R = 7,027$ km and $S = 5.858 \times 10^{40}$ g.cm$^2$.s$^{-1}$ into (1.11) yields a value of 40.8 mas.yr$^{-1}$ for $\mathbf{\Omega}$(gm).

From combination of $R = 7,027$ km and $r_0 = 6,378$ km one obtains $R = 1.10 r_0$. From the truncated series expansion of $\mathbf{B}_{eq}$(sphere) in table 1 follows for $R = 1.10 r_0$, that the ratio between the absolute value of $\mathbf{B}_{eq}$(sphere) and the corresponding value from the dipole model is only 0.0211. Therefore, in a first order calculation of $\mathbf{\Omega}$(gm) only the contribution from the polar region has been taken into account, whereas the contribution from the equatorial region will be neglected. So, we replace the integration of (1.11) by

$$\mathbf{\Omega}(\text{gm}) = \frac{G\mathbf{S}}{c^2 R^3}\left\{4\int_0^{\vartheta_m}(3\cos^2\vartheta - 1)d\vartheta \bigg/ \int_0^{2\pi}d\vartheta\right\} = 0.7542 \frac{G\mathbf{S}}{c^2 R^3}, \tag{5.2}$$

where $\vartheta_m$ is the magic angle ($3\cos^2\vartheta_m - 1 = 0$, $\vartheta_m = 54.74°$). Compared with the integral of (1.11), the contributions from the intervals $\vartheta_m < \vartheta < 180° - \vartheta_m$ and $180° + \vartheta_m < \vartheta < 360° - \vartheta_m$ have been neglected. Instead of the standard precession rate $\mathbf{\Omega}$(gm) = 0.5 $G\mathbf{S}/(c^2R^3)$ from (1.11), we now find the higher precession rate $\mathbf{\Omega}$(gm) = 0.7542 $G\mathbf{S}/(c^2R^3)$ of (5.2). Note that the contributions to $\mathbf{\Omega}$(gm) from the polar and excluded intervals in the dipole model possess opposite signs. As a result, the obtained value for $\mathbf{\Omega}$(gm) is a factor of 1.51 higher than the standard value for $\mathbf{\Omega}$(gm), equivalent to 61.5 mas.yr$^{-1}$. Since we have only calculated the gravitomagnetic field $\mathbf{B}$ at angles $\vartheta = 0°$ (the poles) and $\vartheta = 90°$ (the equator), but not at other values of $\vartheta$, the latter figure can be regarded as a maximum. Moreover, the ratio between the absolute value of $\mathbf{B}_{eq}$(sphere) from a complete series expansion for (4.11) and the corresponding value from the dipole model may yield unity value, so that the standard value from (1.11) applies for $\mathbf{\Omega}$(gm).

In May 2011, the final results of the Gravity Probe B experiment for the two orthogonal contributions to the observed precession rate were published by Everitt *et al.*

[24]. Firstly, a geodetic precession rate of 6601.8 ± 18.3 mas.yr$^{-1}$ (1$\sigma$ error, or 68% confidence interval) was reported, to be compared with the predicted 6606.1 mas.yr$^{-1}$. Since we have not analysed the geodetic effect in our work, we will not discuss it further.

Secondly, a frame-dragging or gravitomagnetic precession rate of 37.2 ± 7.2 mas.yr$^{-1}$ (1$\sigma$ error, or 68% confidence interval) was reported. It is noticed that a value of 39.2 mas.yr$^{-1}$ for the standard precession rate $\Omega$(gm) is given, instead of the previously reported value of 40.9 mas.yr$^{-1}$. Above, we extracted a value of 40.8 mas.yr$^{-1}$ from previously published results. It is noticed that the frame-dragging effect has to be separated off from a solar geodetic effect and an effect due to the proper motion of the guide star. The latter contributions amount to 16.2 ± 0.6 and 20.0 ± 0.1 mas.yr$^{-1}$, respectively (see, ref. [24]), resulting into a total measured effect of 75.4 mas.yr$^{-1}$.

A discussion of the involved errors is given by Everitt *et al.* [24] and by Will [25]. Previously, a critical discussion of the Gravity Probe B mission was given by Brumfiel [26]; see ref. [27] for comment.

## 6. THE LAGEOS SATELLITES

Previously, Ciufolini *et al.* [16] have analysed the Lense-Thirring precessions of the orbits of the LAGEOS and LAGEOS 2 satellites with semimajor axes $a_{\text{LAGEOS}}$ = 12,270 km and $a_{\text{LAGEOS 2}}$ = 12,163 km, respectively. The given eccentricities are small and amount to $e_{\text{LAGEOS}} \approx 0.004$ and $e_{\text{LAGEOS 2}} \approx 0.014$, respectively. Ciufolini and Pavlis [18] reported a result of 99 ± 5 per cent of the value of $\overline{\Omega}_{\text{LT}}$ predicted by (1.12), but they allowed for a total error of ± 10 per cent uncertainty to include unknown and not modelled sources of error. However, the accuracy of this result is disputed by several authors. For example, referring to [18], the GP-B team remarked [28]: "In their measurement, the frame-dragging effect needs to be separated by an extremely elaborate modelling process from Newtonian effects more than 10,000,000 times larger than the effect to be measured." (compare with data from ref. [17]). Additional criticism was given and summarized by Iorio [29]. His conservative error estimate is 20-45%.

Using a radius $r_0$ = 6,378 km for the Earth, one obtains $R \approx 1.9\,r_0$. Table 1 shows that for this value of $R$ the ratio $S/D$ between the more accurate, absolute value of $\mathbf{B}_{\text{eq}}$(sphere) and the corresponding value from the dipole model is 0.61. An extrapolation of the contribution from the higher order terms to the r. h. s. of (4.11) raises the latter ratio to about unity value. Therefore, the deviation from the ideal dipole value of $\mathbf{B}_{\text{eq}}$(sphere) will be neglected. Thus, the Lense-Thirring precession of (1.12) may be unaffected by a possible non-ideal dipole gravitomagnetic field at Earth's equator.

## 7. DISCUSSION

In this section electromagnetic analogues of the gravitomagnetic field $\mathbf{B}$(gm), $\mathbf{B}$(em), are shortly discussed. In addition, the nature of $\mathbf{B}$(gm) is considered more in detail.

### 7.1. THE ELECTROMAGNETIC FIELD

Analogous to (1.1), the electromagnetic field $\mathbf{B} = \mathbf{B}$(em) in the stationary case can be obtained from the simplified Maxwell equations (see, e.g., [9, § 26 and § 30])

$$\nabla \times \mathbf{B} = 4\pi c^{-1}\rho_e \mathbf{v} \quad \text{and} \quad \nabla \cdot \mathbf{B} = 0, \tag{7.1}$$

where $\mathbf{v}$ is velocity and $\rho_e$ is the density of a charge element $dq = \rho_e\,dV$. Comparison of (1.1) and (7.1) shows that the factor $-\beta G^{1/2}\rho$ has been replaced by $+\rho_e$. Note that $\mathbf{B} = \mathbf{B}$(em) has the dimension of a magnetic induction field.

Assuming the ideal dipole located at the centre, the following electromagnetic fields **B**(em) can be calculated from (7.1) for a sphere with homogeneous charge density

$$\mathbf{B}_p(\text{sphere}) = +\tfrac{2}{5} c^{-1} Q r_0^2 R^{-3} \boldsymbol{\omega}, \quad \mathbf{B}_c(\text{sphere}) = +c^{-1} Q r_0^{-1} \boldsymbol{\omega} \text{ and}$$
$$\mathbf{B}_{eq}(\text{sphere}) = -\tfrac{1}{5} c^{-1} Q r_0^2 R^{-3} \boldsymbol{\omega}, \tag{7.2}$$

where $Q$ is the total charge of the sphere. The fields $\mathbf{B}_p(\text{sphere})$, $\mathbf{B}_c(\text{sphere})$ and $\mathbf{B}_{eq}(\text{sphere})$ denote the electromagnetic fields at the poles, at the centre and in the limiting case $R \to \infty$, respectively. These fields are completely analogous to their respective gravitomagnetic counterparts of (3.8), (4.14) and (4.15). Comparison of the corresponding fields $\mathbf{B}(\text{gm})$ and $\mathbf{B}(\text{em})$ shows that the following transformation has been carried out

$$-\beta G^{1/2} m \to +Q. \tag{7.3}$$

In order to obtain additional electromagnetic analogues, the same transformation of (7.3) can be applied to equations (1.8), (4.5), (4.8), (4.9) and (4.11).

Furthermore, the electromagnetic analogue $\mathbf{B}_{eq}(\text{sphere})$ for $R = r_0$ deserves special attention. From the ideal dipole model follows

$$\mathbf{B}_{eq}(\text{sphere}) = -\frac{\mathbf{M}(\text{em})}{r_0^3} = -\tfrac{1}{5} c^{-1} Q r_0^{-1} \boldsymbol{\omega}, \tag{7.4}$$

where **M**(em) is the electromagnetic moment of the sphere. Analogous to (4.11), the electromagnetic field $\mathbf{B}_{eq}(\text{sphere})$ can approximately be calculated. Utilizing the transformation (7.3) to the units of table 1 and figure 2, similar results and curves are obtained for the respective electromagnetic analogues. It is possible, however, that the field $\mathbf{B}_{eq}(\text{sphere})$ reduces to (7.4), when a more complete series expansion would be used.

### 7.2. NATURE OF THE GRAVITOMAGNETIC FIELD

Several authors (see, e.g., [2–8]) have introduced a gravitomagnetic field starting from the Einstein equations in the slow motion and weak field approximation. In general, it is implicitly assumed that the nature of the gravitomagnetic field $\mathbf{B} = \mathbf{B}(\text{gm})$ is totally different from its electromagnetic counterpart $\mathbf{B} = \mathbf{B}(\text{em})$. Since numerous definitions of the gravitomagnetic field are mathematically possible, several different definitions for $\mathbf{B}(\text{gm})$ have been given.

Alternatively, it has previously been proposed that the gravitomagnetic field $\mathbf{B}(\text{gm})$ generated by rotating mass and the electromagnetic field $\mathbf{B}(\text{em})$ due to moving charge are equivalent [2, 4, 5, 20, 30]. Gravitational and electromagnetic phenomena are then connected by the magnetic induction field. Therefore, the same dimensions for $\mathbf{B}(\text{gm})$ and $\mathbf{B}(\text{em})$ are chosen. It is stressed, however, that the dependence of $\mathbf{B}(\text{em})$ on distance $R$, for example, does not change by this particular choice (compare with figure 2). We now briefly outline some consequences of this special interpretation of the gravitomagnetic field.

Identification of $\mathbf{B} = \mathbf{B}(\text{gm})$ in (1.4) as a magnetic induction field implies that rotating electrically neutral matter generates a magnetic field. For example, the rotating Earth generates a magnetic field with a magnetic dipole moment $\mathbf{M} = \mathbf{M}(\text{gm})$, according to (1.5). Then, this equation represents the so-called Wilson-Blackett law. It appears that this relation is approximately valid for *many*, *strongly different*, celestial bodies and for some rotating metallic cylinders in the laboratory as well (for a review, see ref. [2] and references therein; for pulsars, see [30]). It is noticed that the observed values of the magnetic dipole moment **M**(obs) and **S** vary over the large interval of about *sixty decades!* The correct order of magnitude of **M**(obs) compared with **M**(gm) from (1.5) is an important reason to propose that $\mathbf{B}(\text{gm})$ from (1.4) is equivalent to the magnetic induction field **B**(em) generated by moving charge. Discrepancies between the values of **B**(gm) predicted by (1.4) and the observed fields **B**(obs) exist, but they may be attributed to interfering effects from electromagnetic origin [2, 30].

Another indication for the proposed equivalence of the fields **B**(gm) and **B**(em) is formed by the existence of low frequency quasi-periodic oscillations (QPOs). They have been observed for a number of accreting millisecond X-ray pulsars, soft gamma repeaters and black holes. Application of our special interpretation of the gravitomagnetic field results in the deduction of four new *gravitomagnetic* precession frequencies, which have been identified with observed low frequency QPOs [20]. Predictions of the proposed model were compared with observed low frequency QPOs of the pulsars SAX J1808.4−3658, XTE J1807−294, IGR J00291+5934 and SGR 1806−20. The results seem to be compatible with the presented model. Moreover, similar results have been obtained for the stellar black hole XTE J1550−564 and the supermassive black hole Sgr A*.

In addition, it has been pointed out [5] previously, that equivalence of fields **B**(gm) and **B**(em) will lead to an alternative result for the standard gravitomagnetic precession rate **Ω**(gm) (or frame-dragging effect) of a gyroscope. A gyroscope subjected to a total field **B**(tot) = **B**(gm) + **B**(em) cannot distinguish between these fields from different origin and **B** = **B**(tot) has to be substituted into (1.9). However, the gyroscopes in the experimental set-up of Gravity Probe B have carefully been shielded against all external magnetic fields [31, 32]. Then, both fields **B**(gm) and **B**(em) from the Earth may be filtered out and a precession rate **Ω** ≈ 0 may be found. Usually, it is assumed, however, that the field **B**(gm) has properties totally different from the magnetic induction field **B**(em) generated by moving charge. In that case a gravitomagnetic precession rate of about 40.8 mas.yr$^{-1}$ is predicted. As has been discussed in section 5, results from the Gravity Probe B experiment largely confirm the last possibility.

Furthermore, the equivalence between **B**(gm) and **B**(em) will also affect the value of the Lense-Thirring precession rate $\overline{\overline{\Omega}}_{LT}$ of (1.12). In this case, in deducing $\overline{\overline{\Omega}}_{LT}$, the field **B**(gm) in (1.9) has again to be replaced by the total field **B**(tot) = **B**(gm) + **B**(em). Otherwise stated, we should use **M**(tot) = **M**(gm) + **M**(em) in the calculation. From (1.5) the (absolute) value of the gravitomagnetic dipole moment $M$(gm) of the Earth can be calculated. Choosing $\beta$ = + 1 and substituting $S$ = 5.858×10$^{40}$ g.cm$^2$.s$^{-1}$ into (1.5) yields $M$(gm) = 2.524×10$^{26}$ G.cm$^3$. The observed magnetic moment $M$(obs) of the Earth, however, is equal to 7.91×10$^{25}$ G.cm$^3$ (see, e.g., [2]). The dimensionless ratio $\beta_{eff}$ ≡ $M$(obs)/$M$(gm) may reflect the strength of the electromagnetic contribution. It amounts to $\beta_{eff}$ = 0.31 in this case (When no electromagnetic contribution would be present, $\beta_{eff}$ would reduce to unity value). Therefore, the electromagnetic contribution leads to a reduction of the total magnetic field **B**(tot). As a result, the Lense-Thirring precession rate $\overline{\overline{\Omega}}_{LT}$ may be reduced by a factor of 0.31 to

$$\overline{\overline{\Omega}}_{LT} = 0.31 \frac{2G\mathbf{S}}{c^2 a^3 (1-e^2)^{3/2}}. \tag{7.5}$$

Thus, in this case the prediction for the Lense-Thirring prediction is about 31 percent of the predicted value of (1.12), instead of the observed value of 99 ± 10 per cent reported by Ciufolini and Pavlis [18].

## 8. CONCLUSIONS

Usually, the gravitomagnetic field **B**(gm) of a spinning sphere with radius $r_0$ is calculated from an ideal gravitomagnetic dipole, located at the centre of the sphere. For a sphere with a homogeneous mass density the gravitomagnetic field calculated from this model is not valid within the sphere and may deviate in the vicinity of the sphere. Utilizing series expansions, expressions for the gravitomagnetic fields at the pole, **B**(gm)

= **B**$_p$(sphere), and at the equator, **B**(gm) = **B**$_{eq}$(sphere), at distance $R$ from the centre are presented. These series expansions can also be applied for the calculation of the corresponding electromagnetic induction fields **B**(em), generated by a rotating sphere with homogeneous charge density.

The polar gravitomagnetic field **B**$_p$(sphere) of (3.8) is calculated and appears to coincide with the prediction (1.7a) from the ideal dipole model. Thus, the field from (3.8) applies to the poles from $R = r_0$ to $R \to \infty$, i.e., from the surface of the sphere to infinity. In case of $R = r_0$ the field of (3.8) reduces to the ideal dipole result (1.8a).

Calculation of the equatorial gravitomagnetic field **B**$_{eq}$(sphere) appears to be much more complicated. A truncated series expansion for **B**$_{eq}$(sphere) is presented in (4.11). As to be expected, in the limiting case $R \to \infty$ this relation (4.11) coincides with the result of the ideal dipole model (compare equations (4.15) and (1.7b)). For $R = r_0$, however, **B**$_{eq}$(sphere) from (4.11) deviates from the ideal dipole result (1.8b) (see also figure 2). A more extended series expansion of (4.11) for $R = r_0$ will shift the value of **B**$_{eq}$(sphere) of the calculated sum of seven terms $S = -0.0128\ \beta c^{-1} G^{\frac{1}{2}} m r_0^{-1} \omega$ from table 1 into the direction of the ideal dipole value $+ 0.2\ \beta c^{-1} G^{\frac{1}{2}} m r_0^{-1} \omega$ of (1.7b). For values of $R \geq 10 r_0$ both values for **B**$_{eq}$(sphere) coincide within 0.3%. Moreover, for values $R$ smaller than $r_0$ the ideal dipole model completely fails; it predicts the wrong sign and magnitude for the field **B**$_{eq}$(sphere), as can be seen from (1.7b) and table 1.

As an illustration, in figure 4 field pattern are given for **B**(gm) from an ideal gravitomagnetic dipole **M**(gm) (compare with Mc Tavish [33] for a figure drawn to scale) and for the more accurate description of **B**(gm), inspired by expression (4.11). Note that within the sphere the field patterns of panels a and b widely differ, whereas they converge for, say $R \geq 3 r_0$. The corresponding electromagnetic fields **B**(em) yield analogous field pattern.

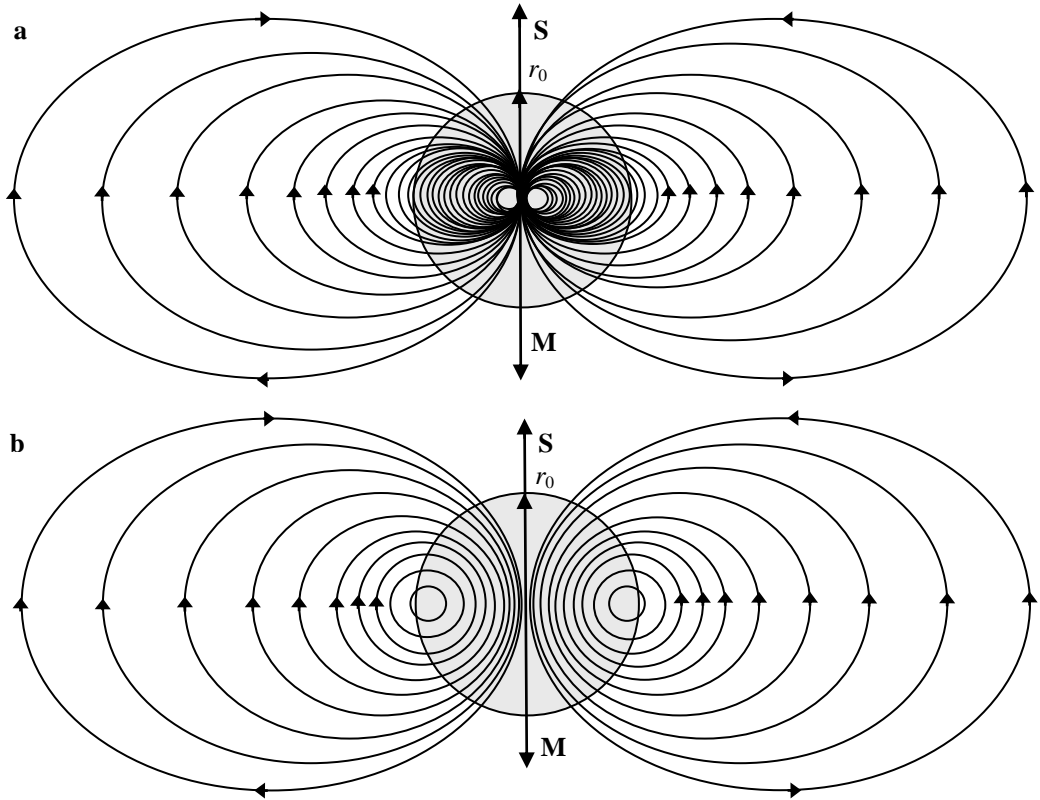

Figure 4. Sketch of the field lines of **B**(gm) from an ideal gravitomagnetic dipole moment **M**, located at the centre of a sphere with radius $r_0$ (panel **a**). A more accurate picture of the field lines of **B**(gm), inspired by relation (4.11), is given in panel **b**.

A discussion of the electromagnetic fields $\mathbf{B}$(em) at the poles, at the centre and at the equator of the sphere, $\mathbf{B}_p$(sphere), $\mathbf{B}_c$(sphere) and $\mathbf{B}_{eq}$(sphere), respectively, has been given in section 7.1.

For the Gravity Probe B experiment with an orbit at $R = 1.10 r_0$ the equatorial gravitomagnetic field $\mathbf{B}_{eq}$(sphere) may lie between zero value and the ideal dipole value $+1/5\ \beta c^{-1} G^{\frac{1}{2}} m r_0^{-1} \omega$ (see section 5). The calculated gravitomagnetic precession rates of a gyroscope, $\mathbf{\Omega}$(gm), are then 61.5 and 40.8 mas.yr$^{-1}$, respectively. Instead of these high values, Everitt *et al.* [24] recently reported a value $\mathbf{\Omega}$(gm) of $37.2 \pm 7.2$ mas.yr$^{-1}$ ($1\sigma$ error, or 68% confidence interval). They deduced a standard result of 39.2 mas.yr$^{-1}$. When gravitomagnetic and electromagnetic field are equivalent (see section 7.2), the predicted value for $\mathbf{\Omega}$(gm) is nearly zero. In table 2 the standard and alternative values of $\mathbf{\Omega}$(gm) for the gyroscopes are compared with reported ones. Fair agreement with the standard prediction is found. See sections 5 and 7.2 for a discussion of these results.

Furthermore, in table 2 the observed and predicted Lense-Thirring precession rates for the LAGEOS/LAGEOS 2 satellites have been compared. All results have been normalized with respect to the standard result of (1.12). The predicted rate for equivalent gravitomagnetic and electromagnetic fields is a factor of 0.33 smaller. All these theoretical results can be compared with the observed value, which differs a factor $0.99 \pm 0.10$ from the standard result. However, the reported error has been criticized by several authors (see section 6).

Table 2. Comparison of observed and theoretical results for Gravity Probe B and the LAGEOS satellites.

| Gravity Probe B | Standard result, based on B(gm) only (mas.yr$^{-1}$) | Observed value (mas.yr$^{-1}$) | Prediction, based on B(gm) + B(em) [f] (mas.yr$^{-1}$) |
|---|---|---|---|
| | 40.8 [a] <br> 39.2 [b] | $37.2 \pm 7.2$ ($1\sigma$) [b] | $\approx 0$ |
| LAGEOS + LAGEOS 2 satellites | Standard result, based on B(gm) only (normalized) [c] | Observed result (normalized) [c] | Prediction, based on B(gm) + B(em) [f] (normalized) |
| | 1 | $0.99 \pm 0.10$ [d] <br> $\pm 0.20\text{-}45$ [e] | 0.33 |

[a] Gravitomagnetic precession rate (or frame-dragging effect), calculated from (1.11). See section 5. [b] Ref. [24].
[c] Lense-Thirring precession from (1.12), normalized to unity value. [d] Ref. [18]. [e] Ref. [29]. [f] See section 7.2.

Summing up, both the results from the Gravity Probe B experiment and from the LAGEOS satellites are in fair agreement with the standard interpretation of general relativity. Perhaps, the proposed equivalence of the gravitomagnetic field and the magnetic field due to moving charge may not be ruled out definitively. Indications of such an equivalence are the approximate validity of the so-called Wilson Blackett law and the compatibility of four observed and predicted low frequency quasi-periodic oscillations (QPOs) of four pulsars and two black holes (see section 7.2). Further observations and analyses will decide between all proposed alternatives.

In any case, the more detailed deduction of the gravitomagnetic field in a sphere may be helpful in numerous future applications.

**ACKNOWLEDGEMENT**

I would like to thank dr. F. Henry-Couannier for a discussion about the ideal dipole character of the gravitomagnetic field near a rotating sphere. The technical help of my son Pieter in publishing this paper is also gratefully acknowledged.